\documentstyle[12pt,epsf]{article}
\begin{document}
\newcommand {\ee}{\end{equation}}
\newcommand {\bea}{\begin{eqnarray}}
\newcommand {\eea}{\end{eqnarray}}
\newcommand {\nn}{\nonumber \\}
\newcommand {\Tr}{{\rm Tr\,}}
\newcommand {\tr}{{\rm tr\,}}
\newcommand {\e}{{\rm e}}
\newcommand {\etal}{{\it et al.}}
\newcommand {\m}{\mu}
\newcommand {\n}{\nu}
\newcommand {\pl}{\partial}
\newcommand {\p} {\phi}
\newcommand {\vp}{\varphi}
\newcommand {\vpc}{\varphi_c}
\newcommand {\al}{\alpha}
\newcommand {\be}{\beta}
\newcommand {\ga}{\gamma}
\newcommand {\Ga}{\Gamma}
\newcommand {\x}{\xi}
\newcommand {\ka}{\kappa}
\newcommand {\la}{\lambda}
\newcommand {\La}{\Lambda}
\newcommand {\si}{\sigma}
\newcommand {\th}{\theta}
\newcommand {\Th}{\Theta}
\newcommand {\om}{\omega}
\newcommand {\Om}{\Omega}
\newcommand {\ep}{\epsilon}
\newcommand {\vep}{\varepsilon}
\newcommand {\na}{\nabla}
\newcommand {\del}  {\delta}
\newcommand {\Del}  {\Delta}
\newcommand {\mn}{{\mu\nu}}
\newcommand {\ls}   {{\lambda\sigma}}
\newcommand {\ab}   {{\alpha\beta}}
\newcommand {\half}{ {\frac{1}{2}} }
\newcommand {\third}{ {\frac{1}{3}} }
\newcommand {\fourth} {\frac{1}{4} }
\newcommand {\sixth} {\frac{1}{6} }
\newcommand {\sqg} {\sqrt{g}}
\newcommand {\fg}  {\sqrt[4]{g}}
\newcommand {\invfg}  {\frac{1}{\sqrt[4]{g}}}
\newcommand {\sqZ} {\sqrt{Z}}
\newcommand {\gbar}{\bar{g}}
\newcommand {\sqk} {\sqrt{\kappa}}
\newcommand {\sqt} {\sqrt{t}}
\newcommand {\reg} {\frac{1}{\epsilon}}
\newcommand {\fpisq} {(4\pi)^2}
\newcommand {\Lcal}{{\cal L}}
\newcommand {\Ocal}{{\cal O}}
\newcommand {\Dcal}{{\cal D}}
\newcommand {\Ncal}{{\cal N}}
\newcommand {\Mcal}{{\cal M}}
\newcommand {\scal}{{\cal s}}
\newcommand {\Dvec}{{\hat D}}   
\newcommand {\dvec}{{\vec d}}
\newcommand {\Evec}{{\vec E}}
\newcommand {\Hvec}{{\vec H}}
\newcommand {\Vvec}{{\vec V}}
\newcommand {\Btil}{{\tilde B}}
\newcommand {\ctil}{{\tilde c}}
\newcommand {\Ftil}{{\tilde F}}
\newcommand {\Stil}{{\tilde S}}
\newcommand {\Ztil}{{\tilde Z}}
\newcommand {\altil}{{\tilde \alpha}}
\newcommand {\betil}{{\tilde \beta}}
\newcommand {\latil}{{\tilde \lambda}}
\newcommand {\ptil}{{\tilde \phi}}
\newcommand {\Ptil}{{\tilde \Phi}}
\newcommand {\natil} {{\tilde \nabla}}
\newcommand {\ttil} {{\tilde t}}
\newcommand {\Rhat}{{\hat R}}
\newcommand {\Shat}{{\hat S}}
\newcommand {\shat}{{\hat s}}
\newcommand {\Dhat}{{\hat D}}   
\newcommand {\Vhat}{{\hat V}}   
\newcommand {\xhat}{{\hat x}}
\newcommand {\Zhat}{{\hat Z}}
\newcommand {\Gahat}{{\hat \Gamma}}
\newcommand {\nah} {{\hat \nabla}}
\newcommand {\gh}  {{\hat g}}
\newcommand {\labar}{{\bar \lambda}}
\newcommand {\cbar}{{\bar c}}
\newcommand {\bbar}{{\bar b}}
\newcommand {\Bbar}{{\bar B}}
\newcommand {\psibar}{{\bar \psi}}
\newcommand {\chibar}{{\bar \chi}}
\newcommand {\bbartil}{{\tilde {\bar b}}}
\newcommand  {\vz}{{v_0}}
\newcommand {\intfx} {{\int d^4x}}
\newcommand {\inttx} {{\int d^2x}}
\newcommand {\change} {\leftrightarrow}
\newcommand {\ra} {\rightarrow}
\newcommand {\larrow} {\leftarrow}
\newcommand {\ul}   {\underline}
\newcommand {\pr}   {{\quad .}}
\newcommand {\com}  {{\quad ,}}
\newcommand {\q}    {\quad}
\newcommand {\qq}   {\quad\quad}
\newcommand {\qqq}   {\quad\quad\quad}
\newcommand {\qqqq}   {\quad\quad\quad\quad}
\newcommand {\qqqqq}   {\quad\quad\quad\quad\quad}
\newcommand {\qqqqqq}   {\quad\quad\quad\quad\quad\quad}
\newcommand {\qqqqqqq}   {\quad\quad\quad\quad\quad\quad\quad}
\newcommand {\lb}    {\linebreak}
\newcommand {\nl}    {\newline}

\newcommand {\vs}[1]  { \vspace*{#1 cm} }

\newcommand {\MPL}  {Mod.Phys.Lett.}
\newcommand {\NP}   {Nucl.Phys.}
\newcommand {\PL}   {Phys.Lett.}
\newcommand {\PR}   {Phys.Rev.}
\newcommand {\PRL}   {Phys.Rev.Lett.}
\newcommand {\CMP}  {Commun.Math.Phys.}
\newcommand {\JMP}  {Jour.Math.Phys.}
\newcommand {\AP}   {Ann.of Phys.}
\newcommand {\PTP}  {Prog.Theor.Phys.}
\newcommand {\NC}   {Nuovo Cim.}
\newcommand {\CQG}  {Class.Quantum.Grav.}


\font\smallr=cmr5
\def\ocirc#1{#1^{^{{\hbox{\smallr\llap{o}}}}}}
\def\ogamma{\ocirc{\gamma}{}}
\def\oM{{\buildrel {\hbox{\smallr{o}}} \over M}}
\def\osigma{\ocirc{\sigma}{}}

\def\overleftrightarrow#1{\vbox{\ialign{##\crcr
 $\leftrightarrow$\crcr\noalign{\kern-1pt\nointerlineskip}
 $\hfil\displaystyle{#1}\hfil$\crcr}}}
\def\overnab{{\overleftrightarrow\nabslash}}

\def\va{{a}}
\def\vb{{b}}
\def\vc{{c}}
\def\tilpsi{{\tilde\psi}}
\def\tbpsi{{\tilde{\bar\psi}}}

\def\Dslash{{}\hbox{\hskip2pt\vtop
 {\baselineskip23pt\hbox{}\vskip-24pt\hbox{/}}
 \hskip-11.5pt $D$}}
\def\nabslash{{}\hbox{\hskip2pt\vtop
 {\baselineskip23pt\hbox{}\vskip-24pt\hbox{/}}
 \hskip-11.5pt $\nabla$}}
\def\xislash{{}\hbox{\hskip2pt\vtop
 {\baselineskip23pt\hbox{}\vskip-24pt\hbox{/}}
 \hskip-11.5pt $\xi$}}
\def\leftnabla{{\overleftarrow\nabla}}

\def\delL{{\delta_{LL}}}
\def\delG{{\delta_{G}}}
\def\delc{{\delta_{cov}}}

\newcommand {\sqxx}  {\sqrt {x^2+1}}   
\newcommand {\gago}  {\gamma_5}
\newcommand {\Ktil}  {{\tilde K}}
\newcommand {\Ltil}  {{\tilde L}}
\newcommand {\Qtil}  {{\tilde Q}}
\newcommand {\Rtil}  {{\tilde R}}
\newcommand {\Kbar}  {{\bar K}}
\newcommand {\Lbar}  {{\bar L}}
\newcommand {\Qbar}  {{\bar Q}}
\newcommand {\Pp}  {P_+}
\newcommand {\Pm}  {P_-}
\newcommand {\GfMp}  {G^{5M}_+}
\newcommand {\GfMpm}  {G^{5M'}_-}
\newcommand {\GfMm}  {G^{5M}_-}
\newcommand {\Omp}  {\Omega_+}    
\newcommand {\Omm}  {\Omega_-}
\def\Aslash{{}\hbox{\hskip2pt\vtop
 {\baselineskip23pt\hbox{}\vskip-24pt\hbox{/}}
 \hskip-11.5pt $A$}}
\def\Rslash{{}\hbox{\hskip2pt\vtop
 {\baselineskip23pt\hbox{}\vskip-24pt\hbox{/}}
 \hskip-11.5pt $R$}}
\def\kslash{
{}\hbox       {\hskip2pt\vtop
                   {\baselineskip23pt\hbox{}\vskip-24pt\hbox{/}}
               \hskip-8.5pt $k$}
           }    
\def\qslash{
{}\hbox       {\hskip2pt\vtop
                   {\baselineskip23pt\hbox{}\vskip-24pt\hbox{/}}
               \hskip-8.5pt $q$}
           }    
\def\dslash{
{}\hbox       {\hskip2pt\vtop
                   {\baselineskip23pt\hbox{}\vskip-24pt\hbox{/}}
               \hskip-8.5pt $\partial$}
           }    
\def\dbslash{{}\hbox{\hskip2pt\vtop
 {\baselineskip23pt\hbox{}\vskip-24pt\hbox{$\backslash$}}
 \hskip-11.5pt $\partial$}}
\def\Kbslash{{}\hbox{\hskip2pt\vtop
 {\baselineskip23pt\hbox{}\vskip-24pt\hbox{$\backslash$}}
 \hskip-11.5pt $K$}}
\def\Ktilbslash{{}\hbox{\hskip2pt\vtop
 {\baselineskip23pt\hbox{}\vskip-24pt\hbox{$\backslash$}}
 \hskip-11.5pt ${\tilde K}$}}
\def\Ltilbslash{{}\hbox{\hskip2pt\vtop
 {\baselineskip23pt\hbox{}\vskip-24pt\hbox{$\backslash$}}
 \hskip-11.5pt ${\tilde L}$}}
\def\Qtilbslash{{}\hbox{\hskip2pt\vtop
 {\baselineskip23pt\hbox{}\vskip-24pt\hbox{$\backslash$}}
 \hskip-11.5pt ${\tilde Q}$}}
\def\Rtilbslash{{}\hbox{\hskip2pt\vtop
 {\baselineskip23pt\hbox{}\vskip-24pt\hbox{$\backslash$}}
 \hskip-11.5pt ${\tilde R}$}}
\def\Kbarbslash{{}\hbox{\hskip2pt\vtop
 {\baselineskip23pt\hbox{}\vskip-24pt\hbox{$\backslash$}}
 \hskip-11.5pt ${\bar K}$}}
\def\Lbarbslash{{}\hbox{\hskip2pt\vtop
 {\baselineskip23pt\hbox{}\vskip-24pt\hbox{$\backslash$}}
 \hskip-11.5pt ${\bar L}$}}
\def\Rbarbslash{{}\hbox{\hskip2pt\vtop
 {\baselineskip23pt\hbox{}\vskip-24pt\hbox{$\backslash$}}
 \hskip-11.5pt ${\bar R}$}}
\def\Qbarbslash{{}\hbox{\hskip2pt\vtop
 {\baselineskip23pt\hbox{}\vskip-24pt\hbox{$\backslash$}}
 \hskip-11.5pt ${\bar Q}$}}
\def\Acalbslash{{}\hbox{\hskip2pt\vtop
 {\baselineskip23pt\hbox{}\vskip-24pt\hbox{$\backslash$}}
 \hskip-11.5pt ${\cal A}$}}

\begin{flushright}
November 2000\\
hep-th/0003275 \\
US-00-01
\end{flushright}

\vspace{0.5cm}

\begin{center}
{\Large\bf 
A Solution of
 the Randall-Sundrum Model\\
and\\
the Mass Hierarchy Problem  }

\vspace{1.5cm}
{\large Shoichi ICHINOSE
          \footnote{
E-mail address:\ ichinose@u-shizuoka-ken.ac.jp
                  }
}
\vspace{1cm}

{\large 
Laboratory of Physics, \\
School of Food and Nutritional Sciences, \\
University of Shizuoka,
Yada 52-1, Shizuoka 422-8526, Japan          
}

\end{center}
\vfill

{\large Abstract}\nl
A solution of the Randall-Sundrum model
for a simplified case (one wall) is obtained. 
It is given by the $1/k^2$-expansion (thin wall
expansion) where $1/k$ is the {\it thickness} of the
domain wall. The vacuum
setting is done by the 5D Higgs potential and the
solution is for a {\it family} of the Higgs parameters.
The mass hierarchy problem is examined.
Some physical quantities in 4D world such as
the Planck mass,
 the cosmological constant, and fermion masses are focussed. 
Similarity to
the domain wall regularization used in the chiral fermion problem
is explained. We examine the possibility that
the 4D massless chiral fermion bound to the domain wall 
in the 5D world can be regarded as the real 4D fermions such as
neutrinos, quarks and other leptons.

\vspace{0.5cm}

PACS NO:\ 04.20.Jb,\ 04.50.+h,\ 11.10.Kk,\ 11.25.Mj,\ 
11.27.+d,\ 11.30.Rd,\ 12.10.Kt\nl


\section{Introduction}
In nature there exists the mass hierarchy such as the Planck mass
($10^{19}$GeV), the GUT scale ($10^{15}$Gev), 
the electro-weak scale ($10^2$GeV),
the neutrino mass($10^{-11}-10^{-9}$GeV) and the cosmological
size($10^{-41}$GeV). How to naturally explain these different 
scales ranging over $10^{60}$ (so huge !) order has been the long-standing
problem (the mass hierarchy problem). One famous approach is the
Dirac's large number theory\cite{PD78}. He tried to explain some
ratios between basic physical quantities (
the electric force/the gravitational force, 
the age of the universe/the period during the light's
passing through the (classical) electron, 
the total mass in the universe/the proton mass)
using the idea of the variable gravitational constant.
Triggered by the development of the string and D-brane theories, 
some interesting new approaches
to the compactification mechanism have recently been proposed
\cite{ADD98,AADD98,RS9905,RS9906} and are applied to the hierarchy problem.
Here we examine the Randall-Sundrum (RS) model which has some attractive
features compared with the Kaluza-Klein compactification.
The model is becoming a strong candidate that could solve
the mass hierarchy problem. 
It looks, however, that the domain wall configuration is
usually introduced "by hand" ( not solving the field equation properly )
and is often "approximated" by some distribution such as
$\delta$-function or $\theta$-function\cite{RS9905,RS9906}. 
Such approximate approach sometimes hinders us from
treating delicate (but important) procedures such as
the boundary condition, the (infrared) regularization and
the cosmological term. 
We present a solution
of the field equation, which clarifies the compactification mechanism
much more than the usual treatment.
Especially the full-fledged treatment of the vacuum
in terms of the 5D Higgs potential is an advantage.
For the purpose of treating the model
starting from the Lagrangian,
we consider the model in a simplified case:\  
One-wall model which was considered in \cite{RS9906}.
An interesting {\it stable} (kink) solution
is found for a {\it family} of vacua. 
The solution does not miss the key points
of the original one. Some new features are\ 
1)\ the (5D) cosmological constant has both the upper bound
and the lower bound (\ref{sol10});\ 2)\ the wall-thickness
parameter $k$ (defined later) is regarded as the most
important quantity to control the whole configuration properly 
and should be bounded both from
below and from above (\ref{chi2});\ 3)\ As a plausible
numerical choice, we propose $k\sim 10^4$GeV to
explain quark, lepton and neutrino masses (Sec.5 and Sec.6).

The domain wall configuration, which is exploited in the RS model,
has been frequently discussed so far in the literature.
Especially the relation between some anomalies is examined in
\cite{CH85}. 
The regularization of the chiral fermion problem
on lattice was examined in
\cite{Kap92,Jan92,Sha93,CH94,Vra98}. 
The similarity to these works
is shown by clarifying the parameters correspondence. 
The difference between them is only the interpretation
of the extra axis;\  In the chiral fermion case it is regarded as
a purely technical axis for the regularization, whereas, in the RS model,
it is a physical axis whose size is too small to measure
at present. The analysis using the RS model
can be regarded as
a geometrical approach to the chiral fermion problem.

At present there exists no sign of the extra dimension(s) experimentally. 
Hence one might wonder about the worth of the present line of research. We 
remind you, however, of the following important aspects behind.
\begin{enumerate}
\item
The higher dimensional view to the present world has been taken
, since Kaluza-Klein\cite{KK21}, by many physicists. The latest
approaches are unified models based on 
the supergravity, the string and the D-brane. This is one basic
standpoint to understand the nature from geometry. The RS model is
one of such approaches and has many distinguished properties
compared with the past ones.
\item
Use of the extra dimension(s) is one possible ingredient, 
independent of the supersymmetry, which can make us
go beyond the standard model.
\item
Many aspects appearing in the RS model overlap with those
in the recent 10-15 years development of the theoretical
physics:
\ the chiral fermion problem (the previous paragraph),
\ anomaly problem,
\ D-brane physics,
\ AdS/CFT,
\ etc..
\item
As reported in ICHEP2000\cite{ICHEP00}, some interesting phenomena await the
experimental tests. 
\end{enumerate}

We introduce the present model in Sec.2. A solution is obtained in Sec.3,
where it is essentially described by three parameters specifying
the Higgs vacuum. In Sec.4, the asymptotic behaviors, in the dimensional
reduction from 5D to 4D, are evaluated for 
the three parameters. In Sec.5, the orders of magnitude for the
two quantities, the thickness parameter ($k$) and the 5D Planck mass
($M$), are examined from the information of the 4D Planck mass,
the 4D cosmological constant and the present experimental status 
of the Newton's law. In Sec.6, similarity between the present analysis
and the the chiral fermion problem is pointed out. Using the 4D fermion
(quarks,leptons) masses, we examine the value of the size of the extra
dimension ($r_c$). The three parameters referred above are
precisely determined in Sec.7, 
where two constraints coming from the boundary
condition are solved for the parameters. We conclude and discuss
in Sec.8. 
 
\section{Model set-up}
We start with the 5D gravitational theory, 
where the metric is Lorenzian, 
with the 5D Higgs potential.
\begin{eqnarray}
S[G_{AB},\Phi]=\int d^5X\sqrt{-G} (-\half M^3\Rhat
-\half G^{AB}\pl_A\Phi\pl_B\Phi-V(\Phi))\com\nn
V(\Phi)=\frac{\la}{4}(\Phi^2-{v_0}^2)^2+\La\com
\label{model1}
\end{eqnarray}
where $X^A (A=0,1,2,3,4)$ is the 5D coordinates and we also use
the notation $(X^A)\equiv (x^\m,y), \m=0,1,2,3.$
$X^4=y$ is the extra axis which is taken to be a space coordinate.
The signature of the 5D metric $G_{AB}$ is $(-++++)$. 
$\Phi$ is a 5D scalar field, $G=\det G_{AB}$, $\Rhat$ is the
5D Riemannian scalar curvature. $M(>0)$ is the 5D Planck mass
and is regarded as the {\it fundamental scale} of this dimensional reduction
scenario. $V(\Phi)$ is the Higgs potential and serves for preparing
the (classical) vacuum in 5D world. 
See Fig.1.
\begin{figure}
\centerline{\epsfysize=4cm\epsfbox{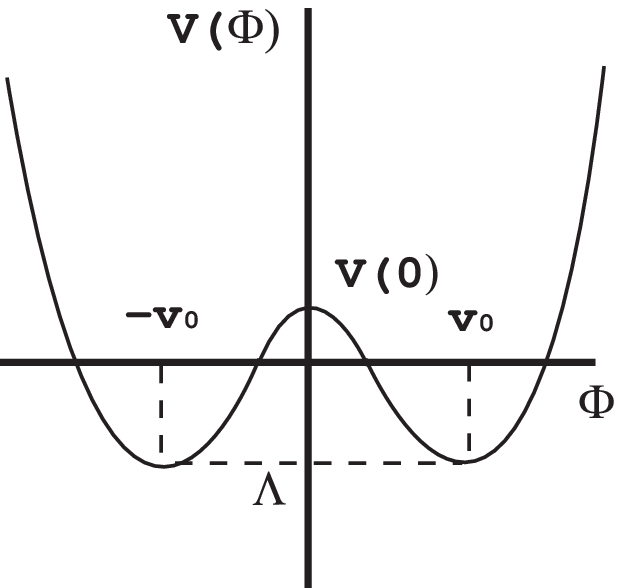}}
   \begin{center}
Fig.1\ The Higgs Potential $V(\Phi)$, (\ref{model1}). 
Horizontal axes: $\Phi$. From (\ref{sol10}), 
$V(0)=\la\vz^4/4+\La>0,\ V(\vz)=\La<0$.
   \end{center}
\end{figure}
The three parameters $\la,\vz$ and $\La$ in $V(\Phi)$ are called here
{\it vacuum parameters}. 
$\la(>0)$ is a coupling, $v_0(>0)$ 
is the Higgs field vacuum expectation value, and $\La$ is the 5D cosmological
constant. 
It is later shown that the sign of $\La$ must be negative for
the proposed domain wall vacuum configuration.
Following \cite{RS9905}, we take the line element shown below.
\begin{eqnarray}
{ds}^2=\e^{-2\si(y)}\eta_\mn dx^\m dx^\n+{dy}^2\com
\label{model2}
\end{eqnarray}
where $\eta_\mn=\mbox{diag}(-1,1,1,1)$. In this choice, the 4D Poincar{\' e}
invariance is preserved. The "warp" factor $\e^{-2\si(y)}$ plays an important
role throughout this paper.
Note that, for the fixed $y$ case ($dy=0$), the metric is the
Weyl transformation of the flat (Minkowski) space $\eta_\mn dx^\m dx^\n$
(See Sec.6).

\section{A solution}
Let us solve the 5D Einstein equation.
\begin{eqnarray}
M^3(\Rhat_{MN}-\half G_{MN}\Rhat )=-\pl_M\Phi\,\pl_N\Phi
+G_{MN}(\half G^{KL}\pl_K\Phi\,\pl_L\Phi+V(\Phi))\com\nn
\na^2\Phi=\frac{\del V}{\del \Phi}\pr
\label{sol1}
\end{eqnarray}
Following Callan and Harvey\cite{CH85},
we consider the case that $\Phi$ depends only on the extra coordinate $y$,
$\Phi=\Phi(y)$.
The above equations reduce to
\begin{eqnarray}
-6M^3(\si')^2=-\half (\Phi')^2+V\com\label{sol2a}\\
3M^3\si''=(\Phi')^2\pr\label{sol2b}
\end{eqnarray}
We note that the "matter equation", the last one of 
(\ref{sol1}), can also be
obtained from $(M,N)=(4,4)$ component of the "gravitational equation",
the first one of (\ref{sol1}) which is given by (\ref{sol2a}).
As the extra space (the fifth dimension), we take 
the real number space ${\bf R}=(-\infty,+\infty)$.
This is a simplified version of the original RS-model\cite{RS9905} 
where $S^1/{\bf Z}_2$ is taken.
We impose the following asymptotic behaviour for the (classical) vacuum 
of $\Phi(y)$.
\begin{eqnarray}
\Phi(y)\ra\pm v_0\com\q
y\ra\pm\infty
\label{sol3}
\end{eqnarray}
This means $\Phi'\ra 0$, and from (\ref{sol2b}), $\si''\ra 0$.
Integrating eq.(\ref{sol2b}), we obtain
\begin{eqnarray}
3M^3\{ \si'|_{y=+\infty}-\si'|_{y=-\infty}\}=
\int^\infty_{-\infty}(\Phi')^2 dy>0\pr\label{sol3b}
\end{eqnarray}
From this result, we are led to
$\si'\ra\pm\om, \si\ra\om |y|$ as $y\ra\pm\infty$, 
where $\om(>0)$ is some
constant to be determined soon. 
We can scale out $M$ in (\ref{sol2a})
by rescaling all fields ($\Phi,\si$), all vacuum parameters ($\la,v_0,\La$)
and the coordinate $y$ with appropriate powers of $M$. 
($
\Phi=M^{3/2}{\tilde \Phi}, \si={\tilde \si}, 
v_0=M^{3/2}{\tilde v_0}, \la=M^{-1}{\tilde \la},
\La=M^{5}{\tilde \La}, X^A=M^{-1}{\tilde X^A},
$)
Therefore we may set $M=1$ without ambiguity. (Only when it is necessary,
we explicitly write down $M$-dependence.)

First we fix the parameter $\om$, by considering $y\ra\pm\infty$
in (\ref{sol2a}), as
\begin{eqnarray}
\om=\sqrt{\frac{-\La}{6}}M^{-\frac{3}{2}}\com
\label{sol4}
\end{eqnarray}
where we see the sign of $\La$ must be {\it negative}, that is,
the 5D geometry must be {\it anti de Sitter} in the
asymptotic regions.

Let us take the following form for $\si'(y)$ and $\Phi(y)$
as a solution.
\begin{eqnarray}
\si'(y)=k\sum_{n=0}^\infty\frac{c_{2n+1}}{(2n+1)!}\{\tanh (ky+l)\}^{2n+1}\com\nn
\Phi(y)=v_0\sum_{n=0}^\infty\frac{d_{2n+1}}{(2n+1)!}\{\tanh (ky+l)\}^{2n+1}\com
\label{sol5}
\end{eqnarray}
where $c$'s and $d$'s are coefficient-constants 
(with respect to $y$) to be determined. 
The free parameter $l$ comes from the {\it translation invariance} of (\ref{sol2a})
and (\ref{sol2b}). A {\it new mass scale} $k(>0)$ is introduced here to make 
the quantity  $k y$ dimensionless. The physical meaning of $1/k$ 
is the "thickness" of the domain wall. 
The parameter $k$, with $M$ and $r_c$(defined later), plays a central role
in this dimensional reduction scenario.
We call $M,k$ and $r_c$ {\it fundamental parameters}.  
The distortion of 5D space-time by the
existence of the domain wall should be small so that 
the quantum effect of 5D gravity
can be ignored and the present {\it classical} analysis is valid. This requires
the condition\cite{RS9905}
\begin{eqnarray}
k\ll M\pr
\label{sol6}
\end{eqnarray}
The coefficient-constants
$c$'s and $d$'s have the following constraints 
\begin{eqnarray}
\sqrt{\frac{-\La}{6}}=k\sum_{n=0}^\infty\frac{c_{2n+1}}{(2n+1)!}
\com\q
1=\sum_{n=0}^\infty\frac{d_{2n+1}}{(2n+1)!}\com
\label{sol7}
\end{eqnarray}
which are obtained by considering the asymptotic behaviours
$y\ra \pm\infty$ in (\ref{sol5}). We will use these constraints
in Sec.7.

We first obtain the recursion relations between
the expansion coefficients,
from the field equations (\ref{sol2a}) and (\ref{sol2b}).
For $n\geq 2$, they are given by 
\begin{eqnarray}
\frac{c_{2n+1}}{(2n)!}-\frac{c_{2n-1}}{(2n-2)!}=\frac{\vz^2}{3}
({D'}_n-2{D'}_{n-1}+{D'}_{n-2})\com\nn
-6C_{n-1}=-\frac{\vz^2}{2}({D'}_n-2{D'}_{n-1}+{D'}_{n-2})
+\frac{\la}{4}\frac{\vz^4}{k^2}(E_{n-2}-2D_{n-1})\ ,
\label{sol8}
\end{eqnarray}
where 
\begin{eqnarray}
D_n=\sum_{m=0}^{n}\frac{d_{2n-2m+1}d_{2m+1}}{(2n-2m+1)!\,(2m+1)!}\com\q
{D'}_n=\sum_{m=0}^{n}\frac{d_{2n-2m+1}d_{2m+1}}{(2n-2m)!\,(2m)!}\ ,\nn
C_n=\sum_{m=0}^{n}\frac{c_{2n-2m+1}c_{2m+1}}{(2n-2m+1)!\,(2m+1)!}\com\q
E_n=\sum_{m=0}^{n}D_{n-m}D_m\pr
\label{sol8b}
\end{eqnarray}
The first few terms, $(c_1,d_1),(c_3,d_3)$, 
are explicitly given as
\begin{eqnarray}
d_1=\pm\frac{\sqrt{2}}{\vz k}\sqrt{\La +\frac{\la \vz^4}{4}}\com\q
c_1=\frac{2}{3 k^2}(\La+\frac{\la \vz^4}{4})\com\nn
\frac{d_3}{d_1}=2+
\frac{1}{k^2}\{ \frac{8}{3}(\La +\frac{\la \vz^4}{4})-\la\vz^2\}
\ ,\ 
\frac{c_3}{c_1}=2+
\frac{1}{k^2}\{ \frac{16}{3}(\La +\frac{\la \vz^4}{4})-2\la\vz^2\}
\ ,
\label{sol9}
\end{eqnarray}
where $\pm$ sign in $d_1$ reflects $\Phi\change -\Phi$ symmetry
in (\ref{sol2a}) and (\ref{sol2b}).
We take the positive one in the following.
We can confirm that the above relations, (\ref{sol8}) and (\ref{sol9}), 
determine all $c$'s and $d$'s
recursively in the order of increasing $n$. 
This is because eq.(\ref{sol8}) is the coupled {\it linear} equation with respect
to ($c_{2n+1},d_{2n+1}$) when lower-order $c'$s and $d'$s are regarded as
obtained quantities. (Note:\ 
$D'_n=\frac{2d_1}{(2n)!}d_{2n+1}+$lower-order terms,\ 
$D_n=\frac{2d_1}{(2n+1)!}d_{2n+1}+$lower-order terms,\ 
$C_n=\frac{2c_1}{(2n+1)!}c_{2n+1}+$lower-order terms,\ 
$E_n=\frac{8{d_1}^3}{(2n+1)!}d_{2n+1}+$lower-order terms. 
)
All coefficients are solved and are described by
the three dimensionless vacuum parameters:\newline 
$(\La +\frac{\la \vz^4}{4})/k^2M^3,\ \la\vz^2/k^2,\ \vz^2/M^3$. 
 
In order for this solution to make sense, as seen from the expression
for $d_1$, the 5D cosmological term $\La$ should be bounded also from below, in addition
to from above.
\begin{eqnarray}
-\frac{\la\vz^4}{4}<\La<0\pr
\label{sol10}
\end{eqnarray}
At this stage the two constraints
(\ref{sol7}) 
are not taken into account. These impose some relations
between vacuum parameters
which will be explained in Sec.7.

\section{Vacuum parameters:\ $M$ and $k$-dependence in the
dimensional reduction}
Let us examine the behaviour of the vacuum parameters
($\La,\vz,\la$) near the 4D world: $k\ra \infty$(the dimensional reduction).
This should be taken consistently with (\ref{sol6}). We will specify
the above limit in the more well-defined way later.
We take the following assumption which will be later checked
using the final solution,
\begin{eqnarray}
\frac{c_{2n+1}}{c_1}\ra O(k^0)\times O(n^0)\com\q
\frac{d_{2n+1}}{d_1}\ra O(k^0)\times O(n^0)\com    \nn
\mbox{as}\q k\ra \infty\ ,\ n\ra \infty \com
\label{asy1}
\end{eqnarray}
where $O(k^0)$ and $O(n^0)$ are some constants of order
$k^0$ and $n^0$. 
$O(n^0)$ behaviour for $n\ra \infty$ is a sufficient condition
for the convergence of the infinite series (\ref{sol5}). Then
the expressions (\ref{sol5}) has the
following asymptotic form, as $k\ra \infty$.
\begin{eqnarray}
\si'(y)\ra kc_1\times\th(ky)\times \mbox{const}
=\frac{2}{3k}(\La+\frac{\la\vz^4}{4})\,\th(ky)\times \mbox{const}\ ,\nn
\Phi(y)\ra\vz d_1\times\th(ky)\times \mbox{const}
=\frac{\sqrt{2}}{k}\sqrt{\La+\frac{\la\vz^4}{4}}\,\th(ky)\times \mbox{const}\ ,
\label{asy2}
\end{eqnarray}
where $\th(y)$ is the step function:\ 
$\th(y)=1$ for $y>0$,$\th(y)=-1$ for $y<0$.
(Note:\ $(\tanh ky)^{2n+1}\ra \th(ky),\ k\ra\infty$.)
Taking relations (\ref{sol3}) and (\ref{sol4}) into account,
(\ref{asy2}) means $\la\vz^4\sim -\La\sim k\sqrt{-\La}\sim k^2\vz^2$. 
These relations say
\begin{eqnarray}
-\La\sim M^3k^2\com\q
\vz\sim M^{3/2}\com\q
\la\sim M^{-3}k^2\q
\mbox{as}\q k\ra \infty\pr
\label{asy3}
\end{eqnarray}
These are {\it leading} behaviour of the vacuum parameters
in the dimensional reduction. 
The first one above is given in the original \cite{RS9905}.
The more precise forms of (\ref{asy3}) will
be obtained, in Sec.7, using the constraints (\ref{sol7}).

\section{Parameter fitting}
In order to express some physical scales in terms of
the fundamental parameters $M$, $k$ and $r_c$(
to be introduced soon), we consider the case that
the 4D geometry is slightly fluctuating around 
the Minkowski (flat) space.
\begin{eqnarray}
{ds}^2=\e^{-2\si(y)}g_\mn (x)dx^\m dx^\n+{dy}^2\com\q
g_\mn=\eta_\mn+h_\mn\com\q h_\mn\sim O(\frac{1}{k})
\pr
\label{para1}
\end{eqnarray}
The leading order $O(k^0)$ results of the previous section remain
valid.

\subsection{The Planck mass}
The gravitational part of 5D action (\ref{model1}) reduces to
4D action as
\begin{eqnarray}
\int d^5X\sqrt{-G}M^3\Rhat\sim
M^3\int_{-r_c}^{r_c}dy\e^{-2\si(y)}\int d^4x\sqrt{-g}R+\cdots\com
\label{para2}
\end{eqnarray}
where the {\it infrared regularization} parameter $r_c$
is introduced. $r_c$ specifies the {\it length} of the extra axis.
Using the asymptotic forms, $\si(y)\sim \om |y|$ as $y\ra \pm\infty$ and
$\om=
\sqrt{\frac{-\La}{6}}M^{-\frac{3}{2}}
\sim k$ as $k\ra \infty$, we can evaluate the order of $M_{pl}$ as
\begin{eqnarray}
{M_{pl}}^2\sim M^3\int_{-r_c}^{r_c}dy\,\e^{-2\om |y|}
=\frac{M^3}{\om}(1-\e^{-2\om r_c})\sim \frac{M^3}{k}\com\label{para3}
\end{eqnarray}
where we have used the {\it 4D reduction condition}:
\begin{eqnarray}
kr_c\gg 1\pr\label{para3b}
\end{eqnarray}
The result (\ref{para3}) is again same as in \cite{RS9905}.
The above condition 
should be interpreted as the precise (well regularized) definition
of $k\ra\infty$ used so far. 
We note $r_c$ dependence in (\ref{para3})
is negligible for $kr_c\gg 1$. This behaviour shows the distinguished
contrast with the Kaluza-Klein reduction (${M_{pl}}^2\sim M^3r_c$)
as stressed in \cite{RS9905}.
\subsection{The cosmological term}
The cosmological part of (\ref{model1}) reduces to 4D action as
\begin{eqnarray}
\int d^5X\sqrt{-G}\La\sim
\La\int_{-r_c}^{r_c}dy\,\e^{-4\si(y)}\int d^4x\sqrt{-g}
\equiv \La_{4d}\int d^4x\sqrt{-g}\com\nn
\La_{4d}\sim 
\frac{\La}{2\om}(1-\e^{-4\om r_c})\sim -M^3 k<0\com\q
kr_c\gg 1\pr\label{para4}
\end{eqnarray}
$\La_{4d}$ is the cosmological term in the 4D space-time. 
It does not, like $M_{pl}$, depend on $r_c$ strongly.
The result says the 4D space-time should also be {\it anti de Sitter}.

\subsection{Numerical fitting}
Let us examine what orders of values should we take for the fundamental
parameters $M$ and $k$. 
( $r_c$ is later fixed by the information of the 4D fermion masses. )
Using the value $M_{pl}\sim 10^{19}$GeV , the "rescaled" cosmological
parameter ${\tilde \La}_{4d}\equiv \La_{4d}/{M_{pl}}^2$ \cite{foot5a}
has the relation:
\begin{eqnarray}
\sqrt{-{\tilde \La}_{4d}}\sim k\sim M^3\times {10}^{-38}\ \mbox{GeV}\com
\label{para5}
\end{eqnarray}
where the relations (\ref{para3}) and (\ref{para4}) are used. 
The unit of $M$ is GeV and
this mass unit is taken in the following. 
The observed value of ${\tilde \La}_{4d}$
is not definite, even for its sign.\cite{Perl99} 
If we take into account the quantum effect, the value of
${\tilde \La}_{4d}$ could run along the renormalization\cite{foot5b}.
Furthermore if we consider
the parameter ${\tilde \La}_{4d}$ represents some "effective"
value averaging over all matter fields, 
the value, no doubt, changes during
the evolution of the universe. (Note the model (\ref{model1})
has no (ordinary) matter fields.) 
Therefore, instead of specifying 
${\tilde \La}_{4d}$, it is useful to consider various possible cases
of ${\tilde \La}_{4d}\sim -k^2$.
Some typical cases are
1) ($k={10}^{-41}, M=0.1$),\ 
2) ($k={10}^{-13}, M={10}^8$)\ 
3) ($k=10, M={10}^{13}$)\ 
4) ($k={10}^4, M={10}^{14}$)\ 
and 
5) ($k={10}^{19}, M={10}^{19}$).
Case 1) gives the most plausible present value of the cosmological constant.
The wall thickness
$1/k={10}^{41}$[GeV$^{-1}$]
, however, is the radius of the present universe.
This implies the extra dimensional effect appears
at the cosmological scale, which should be abandoned.
Case 2) gives  $1/k={10}^{13}$ GeV$^{-1}$ $\sim 1$mm which is
the minimum length at which the Newton's law is checked\cite{ADD98}. 
Usually $k$ should be larger than this value so that
we keep the observed Newton's law.
5) is an extreme case $M=M_{pl}$. 
The fundamental scale $M$ is given by the Planck mass. 
In this case, $r_c\gg 1/k=1/M_{pl}$
is acceptable, while $\sqrt{-{\tilde \La}_{4d}}\sim M_{pl}$
is completely inconsistent with the experiment and
requires explanation. 
Most crucially the condition (\ref{sol6}) breaks down.
Cases 3) and 4) are some intermediate cases which are acceptable except
for the cosmological constant. They will be used in Sec.8. 
At present any choice of ($k,M$)
looks to have some trouble if we take into account the cosmological
constant. We consider the observed cosmological constant
($10^{-41}$GeV) should be explained by some unknown mechanism. 
(No successful explanation of the small cosmological constant exists
\cite{Weinb89}.
Ordinarily (without fine-tuning) 
the quantum-loop correction leads to the case 5)\cite{SI84}.
Compared with case 5), the cases 3) and 4) should be regarded as
"much improved" cases in this respect.)

\section{Domain wall in the chiral fermion problem}
We point out the mechanism presented here has a strong similarity to
that in the chiral fermion determinant.
The {\it interpretation} of the extra axis only is the difference.
The axis is regarded as a real (but hardly measurable) axis here, whereas it is
a regularization axis in the chiral problem. The parameter
correspondence is  
\begin{eqnarray}
\mbox{Randal-Sundrum} &         &\mbox{Chiral Fermion\cite{Vra98,SI98,SI99}}\nn
k:\ {\mbox{(thickness of the wall)}}^{-1} & \leftrightarrow 
& M_F:\ \mbox{1+4 dim fermion mass or}\nn
& &{\mbox{     (thickness of the wall)}}^{-1}\nn
M:\ \mbox{fundamental scale}  & \leftrightarrow & 
1/t:\ \mbox{temperature or}\nn
&& 1/a:\ {\mbox{(lattice spacing)}}^{-1}\nn
r_c:\ \mbox{Infrared reg.}  & \leftrightarrow 
&1/|k^\m|:\ {\mbox{(4D fermion mom.)}}^{-1} \mbox{or}\nn
=\mbox{size of the extra axis}&&  1/m_q:\ {\mbox{(quark mass)}}^{-1}\nn
\label{chi1}
\end{eqnarray}
The condition on $k$ in the RS model, from
(\ref{sol6}) and (\ref{para3b}), is given as
\begin{eqnarray}
\frac{1}{r_c}\ll k \ll M\pr
\label{chi2}
\end{eqnarray}
The corresponding one of the chiral fermion is given by\cite{SI98,SI99}
\begin{eqnarray}
|k^\m|\ll M_F \ll \frac{1}{t}\pr
\label{chi3}
\end{eqnarray}
Both conditions guarantee the mechanism effectively works.

It is known, in the lattice chiral fermion, the choice of the
parameter $M_F$ is so important to produce a good numerical
output, say, the pion mass(\cite{Colum0007} where $M_F$ is denoted
as $M_5$ and is called "domain wall height"). 
Only for well-chosen value of $M_F$, the chiral properties are
controlled. 
In this analogy
the thickness parameter $k$, in the RS model, is considered to be
a key quantity for controlling the whole configuration and 
for fitting with the real world quatities such as
fermion masses. 

The line element of (\ref{model2}) or (\ref{para1})
for a fixed $y$ is the Weyl scaling $g_\mn (x)\ra \e^{-2\si(y)}g_\mn (x)$
of the 4D world:\ $(ds^2)_{4D}=g_\mn(x)dx^\m dx^\n$.
$\si(y)$ is related to the 4D dynamics through the 5D geometrical setting.
The extra dimension $y$ plays the role of the {\it scaling} parameter.  
On the other hand, in the chiral problem, the extra axis can be regarded
as the Schwinger's proper time (inverse temperature) $t$\cite{Sch51}
through the relation \cite{SI98,SI99}:
\begin{eqnarray}
(\frac{\pl}{\pl t}+\Dhat)G(x,y;t)=0\com\q
G(x,y;t)=<x|\e^{-t\Dhat}|y>
\com
\label{chi4}
\end{eqnarray}
where $\Dhat$ is the general 4D operator and $G(x,y;t)$ is the
density matrix. Formally it says 
$\frac{\pl G}{\pl t}\cdot G^{-1}=\frac{\pl}{\pl t}\ln G=-\Dhat$.
This shows the {\it scaling} property of $\ln G$ along the coordinate $t$.
These similar roles of $y$ and $t$ strongly indicate the both mechanisms
are essentially the same. 

In the view of \cite{SI98,SI99}, the "direction" of the system evolvement
of the present model
is given by the sign change of the 5D Higgs field around the origin $y=0$.

As in the Callan and Harvey's paper\cite{CH85}, we can
have the 4D {\it massless chiral} fermion bound to the wall
by introducing 5D {\it Dirac} fermion $\psi$ into (\ref{model1}).
\begin{eqnarray}
S[G_{AB},\Phi]+\int d^5X\sqrt{-G}(\psibar\nabslash\psi+g\Phi\psibar\psi)
\pr
\label{conc1}
\end{eqnarray}
If we {\it regulate} the extra axis by the finite range $-r_c\leq y\leq r_c$,
the 4D fermion is expected to have a small mass 
$m_f\sim k\e^{-kr_c}$ 
(This is known for the two-walls case in \cite{Sha93,Vra98}).
If we take the case 3) in Subsec.5.3 ($k=10,M={10}^{13}$)
and regard the 4D fermion as a neutrino ($m_\nu\sim {10}^{-11}-{10}^{-9}\mbox{GeV}$),
we obtain $r_c=2.76-2.30 \mbox{GeV}^{-1}$. 
If we take case 4) ($k={10}^4,M={10}^{14}$), we obtain
$r_c=(3.45-2.99)\times {10}^{-3}\mbox{GeV}^{-1}$. 
When the quarks or other leptons ($m_q,m_l\sim 10^{-3}-10^2\mbox{GeV}$) are taken
as the 4D fermion, and take the case 4) in Subsec.5.3,
we obtain $r_c=(1.61-0.461)\times {10}^{-3}$GeV$^{-1}$.
It is quite a fascinating idea
to {\it identify the chiral fermion zero mode bound to the wall
with the neutrinos, quarks or other leptons}.

\section{Precise form of vacuum parameters}
As shown in (\ref{asy3}), an interesting aspect of the present
solution is that some family of vacua is selected as the
consistent (classical) configuration. Let us determine the precise
form of (\ref{asy3}) using the two constraints (\ref{sol7}).
In terms of new parameters 
$\Om\equiv \La+\frac{\la}{4}\vz^4 (0<\Om<\frac{\tau}{4}\vz^2), 
\tau\equiv \la\vz^2$, instead of $\La$ and $\la$, the precise
forms are obtained by the $\frac{1}{k^2}$-expansion
for the case $kr_c\gg 1$ as
\begin{eqnarray}
\Om=M^3k^2(\al_0+\frac{\al_1}{(kr_c)^2}+\cdots)
=M^3 k^2\sum_{n=0}^{\infty}\frac{\al_n}{(kr_c)^{2n}}\com\q\nn
\tau=k^2(\ga_0+\frac{\ga_1}{(kr_c)^2}+\cdots)
=k^2\sum_{n=0}^{\infty}\frac{\ga_n}{(kr_c)^{2n}}\com\q\nn
\vz=M^{3/2}(\be_0+\frac{\be_1}{(kr_c)^2}+\cdots)
=M^{3/2}\sum_{n=0}^{\infty}\frac{\be_n}{(kr_c)^{2n}}
\com
\label{vac1}
\end{eqnarray}
where $\al$'s,$\ga$'s and $\be$'s are some numerical (real) numbers
to be consistently chosen using (\ref{sol7}).
If we assume the relation (\ref{asy1}), the infinite series
of (\ref{sol7})  can be safely truncated
at the first few terms. 
In order to demonstrate how the vacuum parameters are fixed, 
we take into account up to $n=2$ in (\ref{sol7}) and 
up to $O(1/(kr_c)^{2\times 2})$ in (\ref{vac1}). 
For general $M, k, r_c$ except 
the condition $kr_c\gg 1$, the coefficients are
determined as
\begin{eqnarray}
\mbox{Vacuum 1:}\ (\al_0,\al_1,\al_2)\equiv (1,0,0)\ \mbox{input}
\qqq\qqq\qqq\nn
(\be_0,\be_1,\be_2;\ga_0,\ga_1,\ga_2)=
(1.6,0,0;4.2,0,0)\com\nn
\mbox{Vacuum 2:}\ (\al_0,\al_1,\al_2)\equiv (1,1,0)\ \mbox{input}
\qqq\qqq\qqq\nn
(\be_0,\be_1,\be_2;\ga_0,\ga_1,\ga_2)=
(1.6,1.1,-0.67;4.2,1.8,1.7)\com\nn
\mbox{Vacuum 3:}\ (\al_0,\al_1,\al_2)\equiv (1,1,1)\ \mbox{input}
\qqq\qqq\qqq\nn
(\be_0,\be_1,\be_2;\ga_0,\ga_1,\ga_2)=
(1.6,1.1,0.45;4.2,1.8,3.5)\pr
\label{vac2}
\end{eqnarray}
We notice our solution
has one {\it free parameter} for each $n$-th set ($\al_n,\be_n,\ga_n$).
This is because the number of constraints for $c'$s and $d'$s is two (\ref{sol7}),
whereas that of quantities to be determined is three (\ref{vac1}). 
Using this freedom we can adjust one of the three vacuum parameters
in the way the observed physical values are explained.
In (\ref{vac2}), we take $\al$'s as the input.
Taking the value $kr_c=10$, Vac.3 has the vacuum expectation value
$v_0M^{-3/2}=1.6$, the cosmological constant
$\La k^{-2}M^{-3}=-1.7$ and the coupling $\la k^{-2}M^3=1.6$.
Other vacua have almost the same values because
the first order term dominate in (\ref{vac1}) for
the thin wall case $kr_c\gg 1$. 
We notice the dimensionless vacuum parameters, 
$v_0M^{-3/2},\ \La k^{-2}M^{-3}$ and $\la k^{-2}M^3$, 
are specified only by the value of $kr_c$ and the input data, say, $\al'$s.
If we specify $k$ and $M$, as considered in Subsec 5.3, the values $v_0,\ \La$
and $\la$ are obtained.
Any higher-order, in principle, can be obtained by
the $\frac{1}{k^2}$-expansion.
 
For the Vac.3, we plot
$\Phi(y)$ and $\Phi'(y)$ for two cases
$kr_c=10$ and $kr_c=20$ in Fig.2.

\section{Discussion and conclusion}
The assumption used in the present explanation is (\ref{asy1}) only.
The first some values in the series of $n$:\ 
$(c_3/c_1,d_3/d_1)=(-1.05,0.48),(c_5/c_1,d_5/d_1)=(-11.9,-0.90)$ 
for Vac.3 with $kr_c=10$ 
indicate its validity. Another evidence of the (strong) convergence
is the fact that the normalization in Fig.2 is quite correctly reproduced.

If we take the boundary condition:\ 
$\Phi(y)\ra\mp v_0\ ,\ y\ra\pm\infty$, instead of (\ref{sol3}),
the opposite {\it chirality} solution is obtained. Both of the pair,
$+$ and $-$ chiralities, are indispensable when the "vector-like"
or non-chiral theory, such as QCD, is taken into account.

One of the fundamental parameters, $r_c$, is introduced in Sec.5 and 6
as the infrared regularization.
This is quite natural in the standpoint of the discretized approach such as lattice.
The treatment, however, should be regarded as an "effective" approach
or a "temporary" stage of the unknown right treatment.
The scale $r_c$ should be introduced naturally in the continuum approach.
If we can generalize the present analysis
to the case of the $S^1/Z_2$ extra space (two-walls case), $r_c$ is interpreted as
the distance between the two walls \cite{RS9905}.
Another interesting possibility is that the scale $r_c$ could be given by
some (at present) unknown mechanism in the space-time manifold such as
the non-commutative geometry\cite{SW99}. It also looks that the AdS/CFT
view\cite{Mald98} of the present model could give a clue to the problem. 

An important task to establish the RS scenario is to introduce
the standard electro-weak model (chiral), QCD (non-chiral) 
and SUSY theories into this scheme.
Recently the bulk standard model has been examined by \cite{CHNOY99}.
In \cite{CG99} a supersymmetric extension is examined.
The RS model has given us richer possibilities for the mass hierarchy
problem than before. It is hoped that the future experiments
can select them.

\begin{figure}
\centerline{\epsfysize=4cm\epsfbox{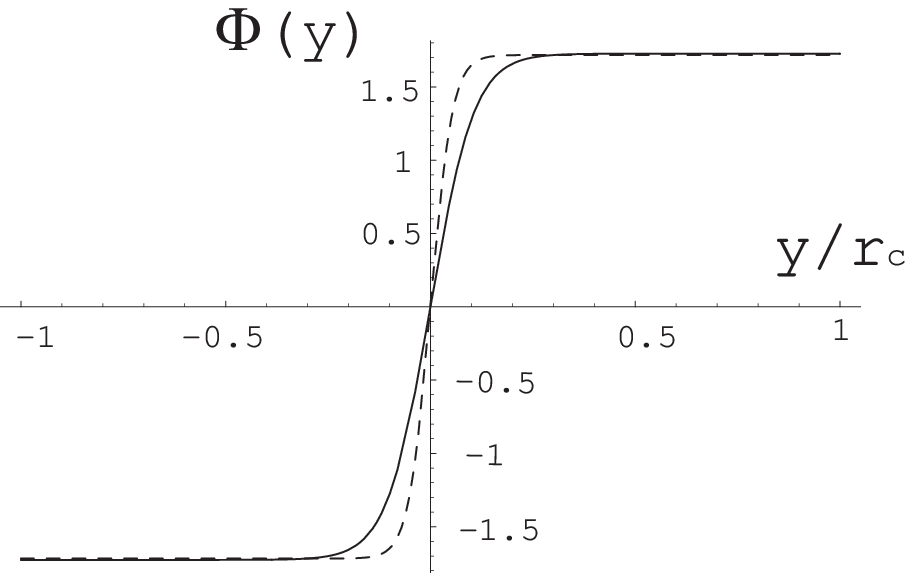}\\
            \epsfysize=4cm\epsfbox{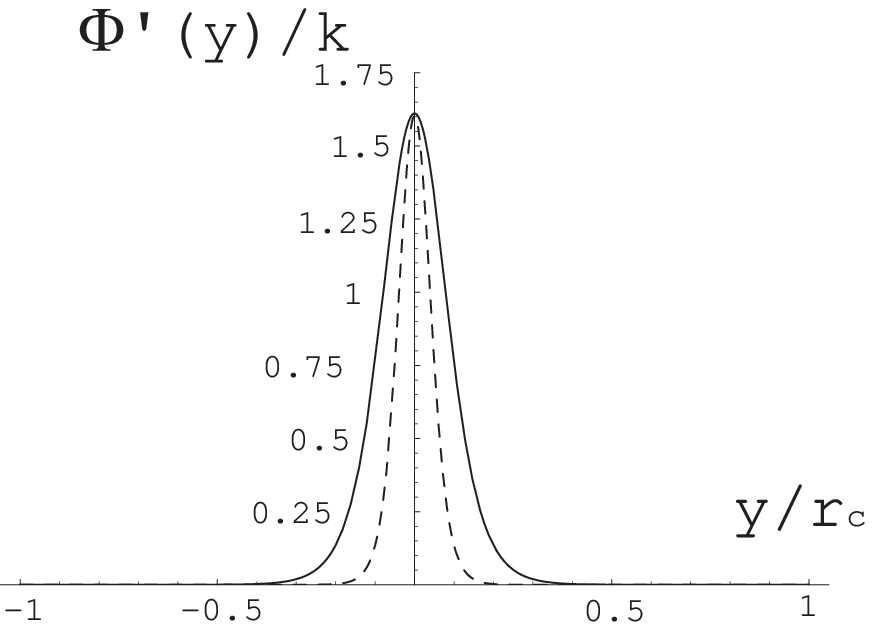}}
   \begin{center}
Fig.2\ 
Vertical axes:\ 5D Higgs field ($\Phi(y)$) and its derivative ($\Phi'(y)/k$);
Horizontal axes: $y/r_c$;
Vacuum 3 of (\ref{vac2});
Solid line: $kr_c=10$,\ Dotted line: $kr_c=20$. 
   \end{center}
\end{figure}

\vspace{2cm}

\vs 1
\begin{flushleft}
{\bf Acknowledgment}
\end{flushleft}
The author thanks G.W.Gibbons for much discussions about the RS-model.
Especially eq.(\ref{sol3b}) was pointed out by him. 
The author also thanks M.Yamaguchi for some comment.

\vs 1



\begin{thebibliography}{99}
\bibitem{foot5a} 
Use of $\La_{4d}$ or ${\tilde \La}_{4d}$ is purely a convension.
\bibitem{foot5b} 
This was shown in \cite{SI95} for the 2 dim $R^2$ gravity
which is renormalizable.
\end{thebibliography}
\end{document}